\documentstyle[preprint,aps]{revtex}
\newcommand{\bmp}{{\mbox{\boldmath $p$}}}
\newcommand{\bms}{{\mbox{\boldmath $s$}}}

\newcommand{\bmr}{{\mbox{\boldmath $r$}}}
\newcommand{\bmz}{{\mbox{\boldmath $r_1$}}}
\newcommand{\bmzp}{{\mbox{\boldmath $r_1'$}}}

\newcommand{\bmq}{{\mbox{\boldmath $q$}}}
\newcommand{\bmb}{{\mbox{\boldmath $b$}}}
\begin{document}
\preprint {WIS-95/45/Sep-PH}
\draft
\date{\today}
\title{Effect of $NN$ correlations on predictions of nuclear
transparencies for protons, knocked-out in high $Q^2$ $(e,e'p)$ reactions}
\author{A. S. Rinat and M.F. Taragin}
\address{Department of Particle Physics, Weizmann Institute of
         Science, Rehovot 76100, Israel}
\maketitle
\begin{abstract}
We study the transparency $\cal T$ of nuclei for nucleons knocked-out
in high-energy semi-inclusive $(e,e'p)$ reactions, using an improved
theoretical input, discussed by Nikolaev et al. We establish that neglect
of $NN$-correlations between
the knocked-out  and core nucleons reduces nuclear transparencies
by $\approx 15 \%$ for light, to $\approx 10\%$ for heavy nuclei.
About the same is predicted for transparencies, integrated over
the transverse or longitudinal momentum of the outgoing proton. Hadron
dynamics predicts a roughly constant $\cal T$ beyond $Q^2\approx $2 GeV$^2$,
whereas for all targets  the largest measured data point
$Q^2$=6.7 GeV$^2$  appears to lie above that plateau. Large error bars on
those data-points preclude a conclusion regarding the onset of
colour transparency.

\end{abstract}
\vspace{1cm}
\par
pacs{: 25.30 Fj, 24.10Eq}
\newpage

\section{Introduction.}
In the past few years  there  has been  a keen  interest in high-energy
semi-inclusive (SI) $A(e,e'p)X$ reactions, from which one wishes to extract
the transparency  ${\cal  T}$ of nuclei
for a knocked-out  nucleon. Special interest centers on the possibility
that from some  threshold on, and initially growing with $Q^2$,
the  medium is far more transparent than  hadronic
dynamics predicts. QCD in fact predicts such behaviour. Depending on the
circumstances, it allocates to the state of a knocked-out proton during
the time it  passes a  nucleus, a small-sized  non-hadronic component. For it
the medium  has an  anomalously  large (colour) transparency (CT) \cite{ct}.

The recently published SLAC NE18 experiment with electron 4-momentum transfers
$1\le Q^2({\rm GeV}^2)\le 6.7$  on various targets \cite{na18,na181}, has
been considered to be a possible testing ground for the onset of
CT. An obvious  prerequisite
for its detection is a demonstration that standard hadronic
dynamics is not capable to account for the data. This is only
possible if both available data and predictions are accurate.

The above explains  maybe the veritable plethora of published
predictions  which  not infrequently refer to  different dynamic
input elements or kinematic details
\cite{ko,mm,n1,n2,n3,rj,om,ffw,rt,fra,cl1}.
Moreover, while  data have been taken and analyzed,
there had been no definitive information available on experimental
details such as cuts on missing mass and momenta, or detector acceptances.
This rendered uncertain, which  calculated transparency should
ultimately be compared with data
and invited tests of the sensitivity  of predicted transparencies for
extreme $'$theoretical$'$ cuts \cite{rt}.

The present note is an outgrow of work by Nikolaev and
co-workers, especially Refs. \cite{n1,n2} and also of \cite{rt}. In all, one
considers SI reactions where the projectile excites high-lying
core states. Application of closure over those states leads to the SI
coincidence cross section, integrated over the electron energy loss.
The corresponding transparency will for brevity be called
the energy integrated one, ${\cal T}^E$.

In general one considers situations where cuts have been applied, i.e. when
the underlying cross sections are integrated over the
entire proton momentum, or over
some component of it. The corresponding transparencies will be called
$'$momentum integrated transparencies$'$, to be
denoted by ${\cal T}^{E,P}$. Those
have been studied by Nikolaev et al in \cite{n1} ${\cal T}^{E,P}$ who report
small effects if $NN$-correlations are retained.

In addition to ${\cal T}^E$,
the transparencies ${\cal T}^{E,P_z}, {\cal T}^{E,P_{\perp}}$  which are
integrated over components of the proton momentum have been
studied in \cite{n2}.
Partly on  the basis of results in  Ref. \cite{n1} for the different
${\cal T}^{E,P}$, $NN$-correlations have
from the onset been disregarded in \cite{n2}.
A comparison of data with predictions for finite  momentum cuts
can be found in \cite{n3}.

The evaluation of the various transparencies discussed  in \cite{n2} is
the cleanest thus far. However, the judgment that
$NN$-correlations may be disregarded seems  to be at variance with other
estimates  on ${\cal T}^E$ \cite{rj,ffw,rt}  and on ${\cal T}^{E,P}$ \cite{pp}.
The same is reported in \cite{rj,fra} for low-energy loss SI reactions, not
permitting closure. Additional material may be found in \cite{bia1}.

There is thus still lacking an actual calculation of the effect of
$NN$-correlations in the otherwise satisfactory description of Nikolaev et
al in \cite{n2}. In the present note we attempt a numerical assessment of their
relevance. In the end we
compare predictions with the NA18 data and discuss the possibility  that the
latter show the onset of colour transparency.

\section{Hadron dynamics for unrestricted and partially restricted
transparencies.}

Without re-derivations we shall in the following borrow results from \cite{rt},
(from here on denoted by I) and cite equations by their number in I. We
start from the assumption that the  $eN$-cross section can be
factored out from the SI cross section and focus on the remaining
SI response or structure function $S^{SI}$. The latter
contains inelastic form factors, connecting the target ground state
to excited of a highly excited core and a high-energy
proton, scattered in the field of the core (I.1).

A considerable simplification results, if the  excitation of high core
states permits closure over those to be performed.
The result  for the SI response per nucleon  is (cf. I.3)
\begin{mathletters}
\label{a1}
\begin{eqnarray}
S^{SI}(q,\omega,\bmp)
&=&\delta(\omega+\langle\Delta\rangle-e_{p})
\int \int d{\bmr}_1 d{\bmr}'_1 e^{ i\bmp'({\bmr}_1-{\bmr}'_1)}
\rho_1({\bmr}_1,{\bmr}_1') \tilde R(q,{\bmr}_1,{\bmr}_1')
\label{a1a}\\
\tilde R(q,\bmr_1,\bmr_1')&=& \bigg ( \prod_j \int d{\bmr}_j\bigg )
\frac{\rho_A({\bmr}_1,{\bmr}_1';{\bmr}_j)}{\rho_1({\bmr}_1,{\bmr}_1')}
\prod_{l\ge 2}
\bigg \lbrack 1+\gamma(q,{\bmr}_1-{\bmr}_l;{\bmr_1'}-\bmr_l)\bigg \rbrack
\label{a1b}\\
&=& \bigg ( \prod_j \int d{\bmr}_j\bigg )
\frac{\rho_A({\bmr}_1,{\bmr}_1';{\bmr}_j)}{\rho_1({\bmr}_1,{\bmr}_1')}
\bigg \lbrack 1+ \Sigma_{l\ge 2}
\gamma(q,{\bmr}_1-{\bmr}_l;{\bmr_1'}-\bmr_l)+.....
\bigg \rbrack
\label{a1c}
\end{eqnarray}
\end{mathletters}
The above SI response depends parametrically on
the momentum-energy loss  $(q,\omega)$ of the incident electron,
the momentum and total energy $\bmp, e_{\bmp}$ of the knocked-out
proton and on some average excitation energy $\langle\Delta\rangle$.
The momentum of the struck proton (minus the missing momentum)
is $\bmp'=\bmp-\bmq$. The following dynamical input elements are required
for an evaluation of (1):

1) $n$-particle density matrices $\rho_n(\bmr_1,\bmr_1';\bmr_j)$, diagonal
in all coordinates except in the one of the struck particle, which has
arbitrarily been chosen to be 1.

2) The operator $\gamma$ which describes the
FSI factor $\tilde R$ in (1b), and which in turn is
governed by the scattering of the knocked-out proton with the core.
Eq. (\ref{a1c}) is its multiple-scattering expansion of (1b) in terms of
the off-shell scattering of '1' from  individual core nucleons.

In view of the large momentum transfers imparted onto the knocked-out nucleon,
its scattering from the core is conveniently described by Glauber theory
(cf. \cite{mg}. By way of illustration we give the result for
$\gamma$ in Eq. (\ref{a1}) for scattering on an isolated core-nucleon at
the origin  ($\bmr=\bmb,z$)
\begin{mathletters}
\label{a2}
\begin{eqnarray}
\gamma(q,\bmr,\bmr')
&=&\bigg (1+\Gamma_q^{off}(\bmb,z)\bigg )\bigg(1+
\lbrack\Gamma_q^{off}(\bmb',z')\rbrack^*\bigg )-1
\label{a2a}\\
&=& \Gamma_q^{off}(\bmb,z)+\lbrack\Gamma_q^{off}(\bmb',z')\rbrack^*
+\Gamma_q^{off}(\bmb,z)\lbrack\Gamma_q^{off}(\bmb',z')\rbrack^*,
\label{a2b}
\end{eqnarray}
\end{mathletters}
The generally off-shell scattering is in the impact representation described
by a corresponding off-shell profile $\Gamma^{off}$. For diffractive $NN$
interactions the off-shell profile may
approximately be related to their on-shell analog $\Gamma=\Gamma^{on}$
\begin{mathletters}
\label{a3}
\begin{eqnarray}
\Gamma_q^{off}(b,z)&\approx&
\theta(-z)\Gamma_q(b)\
\label{a3a}\\
\Gamma&=&e^{i\chi_q(b)}-1\approx
-(\sigma_q^{tot}/2)(1-i\tau_q)A(q,\bmb),
\label{a3b}\\
A(q,\bmb)&\approx& \frac {(Q_q^0)^2}{4\pi}{\rm exp}\lbrack-(bQ_q^0/2)^2
\label{a3c}
\end{eqnarray}
\end{mathletters}
where we used standard
$NN$ scattering parameters:  the  $NN$ total cross  section
$\sigma^{tot}$, the ratio $\tau$ of real and imaginary parts of the
forward $NN$ elastic scattering amplitude and $Q_0^{-1}$,
the range of $\Gamma_q(b)$; zero-range corresponds to
$A(q,\bmb)\to\delta^2(\bmb)$. In the impact parameter representation
$'$partial$'$ inelastic contributions $\sigma^{p,inel}(b)$ to elastic $NN$
scattering are as follows related to the imaginary part of the
eikonal phases $\chi$ in (\ref{a3b})
\begin{eqnarray}
\sigma_q^{p,inel}(b)
&\equiv& 1-e^{-2{\rm Im}\chi_q(b)} \approx A(q,\bmb)\sigma_q^{inel}
\nonumber\\
\sigma_q^{inel}&=&\sigma_q^{tot}-\sigma_q^{el}
\label{a4}
\end{eqnarray}

Since we shall limit ourselves to FSI caused by binary collisions (i.e.
contributions in (1c) up to first order  in $\gamma$) we need only discuss
the two lowest densities. As in I, we parameterize as follows the non-diagonal
single-particle density matrix \cite{vn}
\begin{eqnarray}
\rho_1(\bmz,\bmzp)&\approx&\rho(\mbox{\boldmath $ S$)}\int d\mbox
{\boldmath $ S$}'\rho_1(\bms,\mbox{\boldmath $ S$}')\approx
\rho(\mbox{\boldmath $ r$}_1)
\Sigma(\bms)
\nonumber\\
\Sigma(\bms)&=&\int \frac {d^3\bmp}{(2\pi)^3} n(\bmp)e^{-i\bmp\bms},
\label{a5}
\end{eqnarray}
where ${\mbox{\boldmath $ S$}}=(\bmz +\bmz')/2; \bms=\bmz-\bmz'$.
$\rho(\bmr)=\rho_1(\bmr,\bmr)$ and $n(\bmp)$ are the single
nucleon density and momentum distributions.

More problematic is the half-diagonal $\rho_2$. Only recently have calculations
been made for nuclear matter and some light nuclei \cite{em,do}. For our
purposes suffices the following interpolating approximation \cite{grs}
\begin{eqnarray}
\rho_2(\bmr_1,\bmr_1';{\bmr}_2)
&=& \rho_1(\bmr_1,\bmr_1') \rho({\bmr}_2)
\zeta_2(\bmr_1,\bmr_1';\bmr_2)
\nonumber\\
\zeta_2(\bmr_1,\bmr_1';\bmr_2)
&\approx& \sqrt {g(\bmr_1-\bmr_2) g(\bmr_1'-\bmr_2)},
\label{a6}
\end{eqnarray}
with $g(r)$ the pair distribution function (see \cite{rt1} for some remarks
on the use of (6) for fermions). Using (\ref{a4}) and  (\ref{a5}),
Eqs. (\ref{a1}) become
\begin{mathletters}
\label{a7}
\begin{eqnarray}
S^{SI}(q,\omega,\bmp)&=&\delta(\omega+\langle\Delta\rangle-e_{\bmp})
\int d\bms e^{i\bmp'\bms}\Sigma(|\bms|)
\int d\bmr_1\rho(\bmr_1)
\tilde R(q,\mbox{\boldmath $r_1$},s_z)
\nonumber\\
&=&\delta(\omega+\langle\Delta\rangle-e_{\bmp}) \int d\bms
e^{i\bmp'\bms}\Sigma(|\bms|) {\tilde G}(q,\bms)
\label{a7a}\\
{\tilde G}(q,\bms)&=&\int d{\mbox{\boldmath $r_1$}}\rho(\bmr_1)
\tilde R(q,\mbox{\boldmath $r_1$},\bms)
\label{a7b}
\end{eqnarray}
\end{mathletters}
The above will be confronted with
the standard  Plane Wave Impulse Approximation (PWIA), where by
definition the knocked-out proton exits without being affected by the
medium. The corresponding response is obtained from (\ref{a7}) putting
$\tilde R,\tilde G \to 1$
\begin{eqnarray}
S^{SI,PWIA}(q,\omega,\bmp)=
\delta(\omega+\langle\Delta\rangle-e_{\bmp}) \int d\bms
e^{ i{\bmp'\bms} }\Sigma(|\bms|)\int d\bmr_1\rho(\bmr_1)
=\delta(\omega+\langle\Delta\rangle-e_{\bmp}) n(\bmp')
\label{a8}
\end{eqnarray}
We define ${\cal T}^{SI}$ as the ratio of the experimental
yield and its theoretical PWIA approximation or, equivalently, as the ratio
of the corresponding SI responses. In transparencies, cuts are applied in
the full cross section and in the corresponding PWIA approximation.
Cuts in cross sections should be applied both
{}From  (7), (8) one then obtains for a coincidence experiment
\begin{mathletters}
\label{a9}
\begin{eqnarray}
{\cal T}(q,\omega,\bmp)
&\equiv & \frac{S^{SI}(q,\omega,\bmp)}
{S^{SI,PWIA}(q,\omega,\bmp)}=\frac{ \int d\bms
e^{i\bmp'\bms}\Sigma(|\bms|){\tilde G}(q,\bms)}
{\int d\bms e^{i\bmp'\bms}\Sigma(|\bms|)}
\label{a9a}\\
&=& [n(p')]^{-1}\int d\bms e^{i\bmp'\bms}\Sigma(|\bms|){\tilde G}(q,\bms)
\nonumber\\
&=& [n(p')]^{-1}\int \frac {d\bmp"}{(2\pi)^3} n(\bmp"-\bmp') G(q,\bmp"),
\label{a9b}
\end{eqnarray}
\end{mathletters}
with $G$ the Fourier transform of $\tilde G$.
The absence of $\omega$ in (\ref{a9}) is an artifact of the applied closure.
It causes ${\cal T}\to {\cal T}^E$, with ${\cal T}^E$ the transparency,
when SI cross sections, integrated over the electron energy loss $\omega$
are used.

Eq. (\ref{a9}) is formally like (I.14). The difference lies
in the approximation used in I, used for its evaluation.
Eq. (\ref{a3}) shows, that the
operator $\gamma$ in (1) for the coincidence response,
is composed of two off-shell scattering phases
$e^{i\chi^{off}(q,b,z)}=\Gamma^{off}(q,\bmb,z)+1$ at different ($\bmb,z$).
In I it has been assumed that it is a reasonable approximation to assume
those scatterings to proceed for identical
impact parameters $\bmb$, but different $z$. Nikolaev et al
have since pointed out that this restriction may be relaxed \cite{n2}
and  they
have evaluated (9) in a mean-field approximation for nuclei, disregarding
spatial correlations. However, one may proceed as in I where those have
been retained, keeping as in \cite{n2} different trajectories in $\gamma$,
Eq. (1). Limiting ourselves to binary collisions, one first defines
the following Final State Interaction (FSI) component phases
\begin{mathletters}
\label{a10}
\begin{eqnarray}
\tilde \Omega(q,\bmr_1,\bmr_1')&=&-(A-1)\frac {\sigma^{tot}}{2}(1-i\tau)
\int_{-\infty}^0 dz\int d^2\bmb \,A(b)\rho(\bmr_1-\bmr)
\zeta_2(\bmr;\bmr'-\bms)
\nonumber\\
&\to& -(A-1)\frac {\sigma^{tot}}{2}(1-i\tau)
\int_{-\infty}^0dz\rho(\bmb_1,z-z_1) \zeta_2(0,z;\-\bms_{\perp},z-s_z))
\label{a10a}\\
\tilde \Omega'(q,\bmr_1,\bmr_1')&=&(\Omega(q,\bmr_1',\bmr_1)^*
\nonumber\\
&=&-(A-1)\frac {\sigma^{tot}}{2}(1+i\tau)
\int_{-\infty}^0dz\rho(\bmb_1-\bms_{\perp},z_1-z-s_z)
\zeta_2(0,z;-\bms_{\perp},z+s_z)
\label{a10b}\\
\tilde M(q,\bmr_1,\bms)&=&(A-1) \sigma^{tot\,el}
e^{-{s_{\perp}}^2Q_0^2/8}
\bigg \lbrack \int_{-\infty}^0dz J-\theta(-s_z)\int_{-|s_z|}^0\,dz J
\bigg\rbrack
\nonumber\\
J&=&\rho\bigg (\bmb_1-\bms_{\perp}/2,z_1-z\bigg )
\zeta_2\bigg (\bms_{\perp}/2,z;-\bms_{\perp}/2,z-s_z\bigg ),
\label{a10c}
\end{eqnarray}
\end{mathletters}
where the second equation (\ref{a10a}) is the zero range limit
of the first one. Those define  the FSI factors in the
first cumulant approximation \cite{gr2}
\begin{eqnarray}
\tilde R&=&{\rm exp}[\tilde \Omega+\Omega'+\tilde M]
\nonumber\\
\tilde G&=&\int d\bmr_1 \tilde R(q,\bmr_1,\bmr_1')
\label{a11}
\end{eqnarray}
Substitution of (11) into (9) provides the coincidence SI transparency
${\cal T}^E$ as function of missing  proton momenta.

The neglect of $NN$-correlations amounts to putting $\zeta_2$ in (10), thus
\begin{mathletters}
\label{a100}
\begin{eqnarray}
\tilde {\Omega}_0(q,\bmr_1,\bmr_1')&=&-(A-1)\frac {\sigma^{tot}}{2}(1-i\tau)
\int_{-\infty}^0 dz\int d^2\bmb \,A(b)\rho(\bmr_1-\bmr)
\nonumber\\
&\to& -(A-1)\frac {\sigma^{tot}}{2}(1-i\tau)
\int_{z_1}^{\infty}d{\bar z}\rho(\bmb_1,{\bar z})
\label{a100a}\\
\tilde {\Omega}'_0(q,\bmr_1,\bmr_1')&=&(\Omega_0(q,\bmr_1',\bmr_1)^*
\nonumber\\
&=&-(A-1)\frac {\sigma^{tot}}{2}(1+i\tau)
\int_{z_1}^{\infty}d{\bar z}\rho(\bmb_1-\bms_{\perp},{\bar z}-s_z)
\label{a100b}\\
{\tilde M}_0(q,\bmr_1,\bms)&=&(A-1) \sigma^{tot\,el}
e^{-{s_{\perp}}^2Q_0^2/8}
\bigg \lbrack \int_{-\infty}^0dz J-\theta(-s_z)\int_{-|s_z|}^0\,dz J
\bigg \rbrack
\nonumber\\
J&=&\rho\bigg (\bmb_1-\bms_{\perp}/2,z_1-z\bigg )
\label{a100c}
\end{eqnarray}
\end{mathletters}

In I we have tested several relaxations on the proton momentum in
the above $\cal T$, moving from semi-inclusive
to ever more inclusive reactions. Also here we consider the same when yields
for fixed $q$ are integrated over energy loss as well as the same for
unobserved  proton momentum, transverse to $\hat{\bmq}$, i.e. (cf. I.15)
\begin{eqnarray}
{\cal T}^{E,P_{\perp}}(q,p_z)
\equiv  \frac {\int d^2\bmp_{\perp} S^{SI}(q,\bmp)}
{\int d^2\bmp_{\perp} S^{SI,PWIA}(q,\bmp)}
=\frac {\int ds_z
e^{ip'_zs_z }\Sigma(s_z)\tilde G(q,s_z)}
{ \int d{s_z} e^{ip'_zs_z }\Sigma(s_z)}
\label{a12}
\end{eqnarray}

The same for a maximal $p_z$ cut reads
\begin{eqnarray}
{\cal T}^{E,P_z}(q,\bmp_{\perp})
\equiv  \frac {\int dp_z S^{SI}(q,\bmp)}
{\int dp_z S^{SI,PWIA}(q,\bmp)}
=\frac {\int d\bms_{\perp}
e^{i\bmp'_{\perp}\bms_{\perp} }\Sigma(|\bms_{\perp}|)
\tilde G(q,|\bms_{\perp}|)}
{ \int d{\bms_{\perp}} e^{i\bmp'_{\perp}\bms_{\perp} }\Sigma(|\bms_{\perp}|)}
\label{a13}
\end{eqnarray}

\section{Results and discussion.}

Using the parameters of Table I in I, we have computed transparencies under
varying experimental and theoretical conditions. The discussion here will be
limited to  ${\cal T}^E$, Eqs. (9), and ${\cal T}^{E,P_{\perp}}$,
Eq. (\ref{a12}), and both  use
energy-integrated cross sections. Their difference lies in the
transverse proton momentum $\bmp_{\perp}$ which in the former case is observed
and in the latter is not.

Figs. 1a-c  show  the above mentioned transparencies
for C, Fe and Au for a selected value
$p_z=q$ and as function of the four $Q^2$ values of the
NE18 experiment. In each case we show results with $NN$-correlations included
and with those removed, i.e. using (11) in respectively  (10) and
{\ref{a100}).  In addition we added the result worked out in I and
\cite{rj} for the above mentioned approximation where the
proton-trajectories in $\gamma$, Eq. (1), having equal
impact parameters $\bmb_1$. The following emerges:

1) For the smallest measured $Q^2$=1 GeV$^2$ the equal impact parameter Ansatz
considerably overestimates the proper ${\cal T}^E$.
For all other data points the overestimate is less than 10$\%$, which is not
small on the scale of effects one wishes to investigate.

2) As has also been observed in I, taking a maximal cut in $\bmp_{\perp}$
leads only to minor changes in ${\cal T}$.

3) The inclusion of $NN$-correlations in ${\cal T}^E,{\cal T}^{E,P_{\perp}}$
based on binary collision FSI, enhances both chosen transparencies in a hardly
$Q^2$-independent fashion by $\approx 15\%$ for C to $\approx 10\%$ for Au.

The latter outcome is well-established and numerically significant, yet
no clear-cut conclusions follow from a comparison of data.  Predictions when
$NN$-correlations are included or in their absence are both within error bars,
and from the current data one cannot prove the need for the above correlations.

On the theoretical side there is uncertainty about the influence of higher
order
contributions. For instance Nikolaev et al \cite{n1} estimated the effect of
ternary collisions between the knocked-out proton and a pair of
correlated core nucleons on ${\cal T}^{E,P}$ and claim that those
about halve correlation effects in binary collisions FSI alone.
We have already remarked in I that not all ternary collision terms in (1),
quadratic in $\gamma$ have been retained in \cite{n1}. It is desirable to
include corrections due to FSI from higher order collision, but a reliable
calculation may well be  difficult.

Regarding the data,
with the exception of C, the lowest $Q^2$ data points  agree with
predictions and reflect the relatively low $NN$ cross  sections, which enter
(9) and (\ref{a12}). Their increase for increasing energy to fairly
constant values leads to a predicted plateau in ${\cal T}$.
The reported increase of $\cal T$ for large $Q^2$ falls therefore outside
hadronic predictions. A similar observation with much greater accuracy
may mark the onset of Colour Transparency, but the
large error bars  preclude such a conclusion at the moment.

We conclude by comparing the above formalism for SI processes
and the one used in \cite{rt1} for totally inclusive $A(e,e')X$ ones.
Both emphasize the Final State Interaction of a
struck proton with the core and their descriptions have therefore
common features. It is therefore gratifying to see a similar
measure of agreement between data and predictions for both type of
reactions.


\newpage

{\bf  Figure Captions.}

Fig. 1a.  Transparencies  of  C for  a proton,
knocked out in $C(e,e'p)X$  for NE18 kinematics and $p_z=q$, as function of
$Q^2$. Drawn and dotted lines correspond to ${\cal T}^{E}$, Eq. (9),
with the effect of $NN$-correlations  included, respectively neglected.
Long dashes and dash-dotted curves are  pararle results for
${\cal T}^{E,P_{\perp}}$, Eq. (\ref{a12}). Short dashes are results from I
for (9) including correlations, assuming $\bmb_1'=\bmb_1$ in (1).
Data are from O$'$Neill $et\,al$
\cite{na181} and differ slightly from previously published  data by Makins
$et\,al$ \cite{na18}.

Fig. 1b. Same as Fig. 1a for Fe. Data are from O'Neill $et\,al$
\cite{na181}.

Fig. 1c. Same as Fig. 1a for Au.




\vspace{2cm}
\par

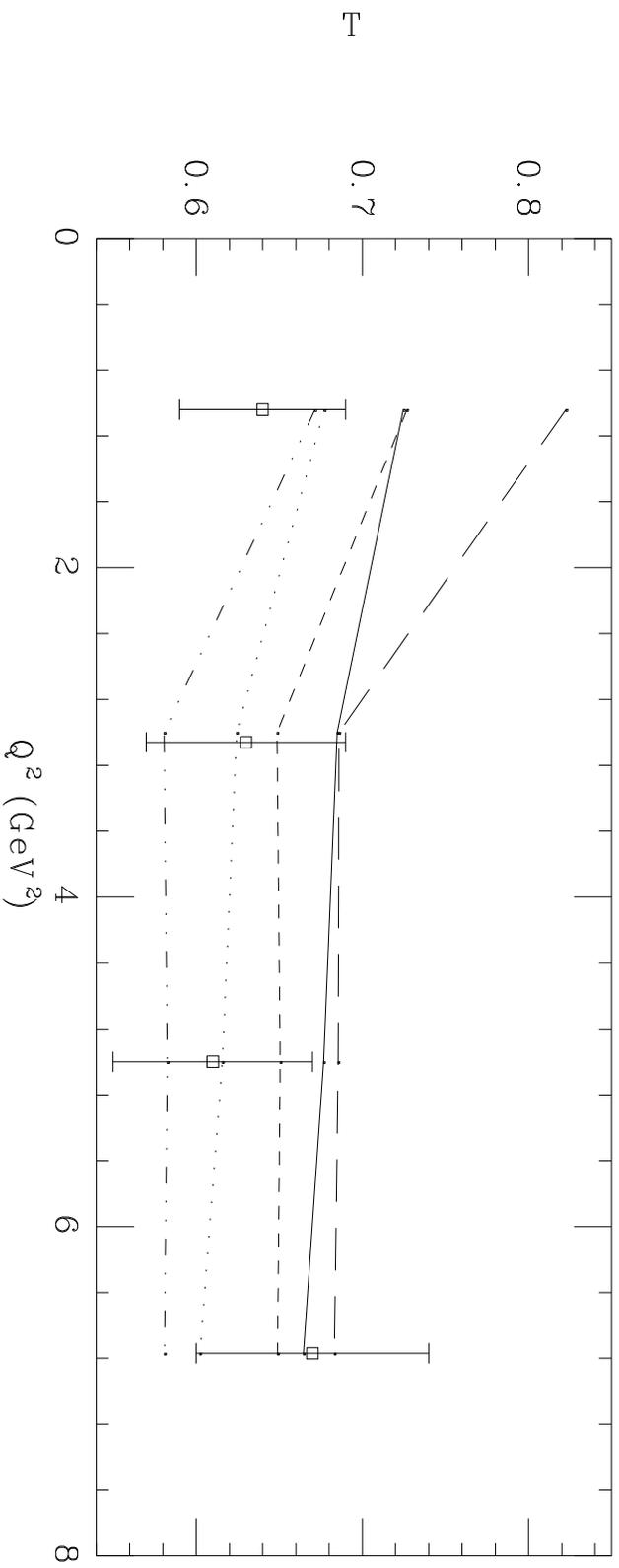
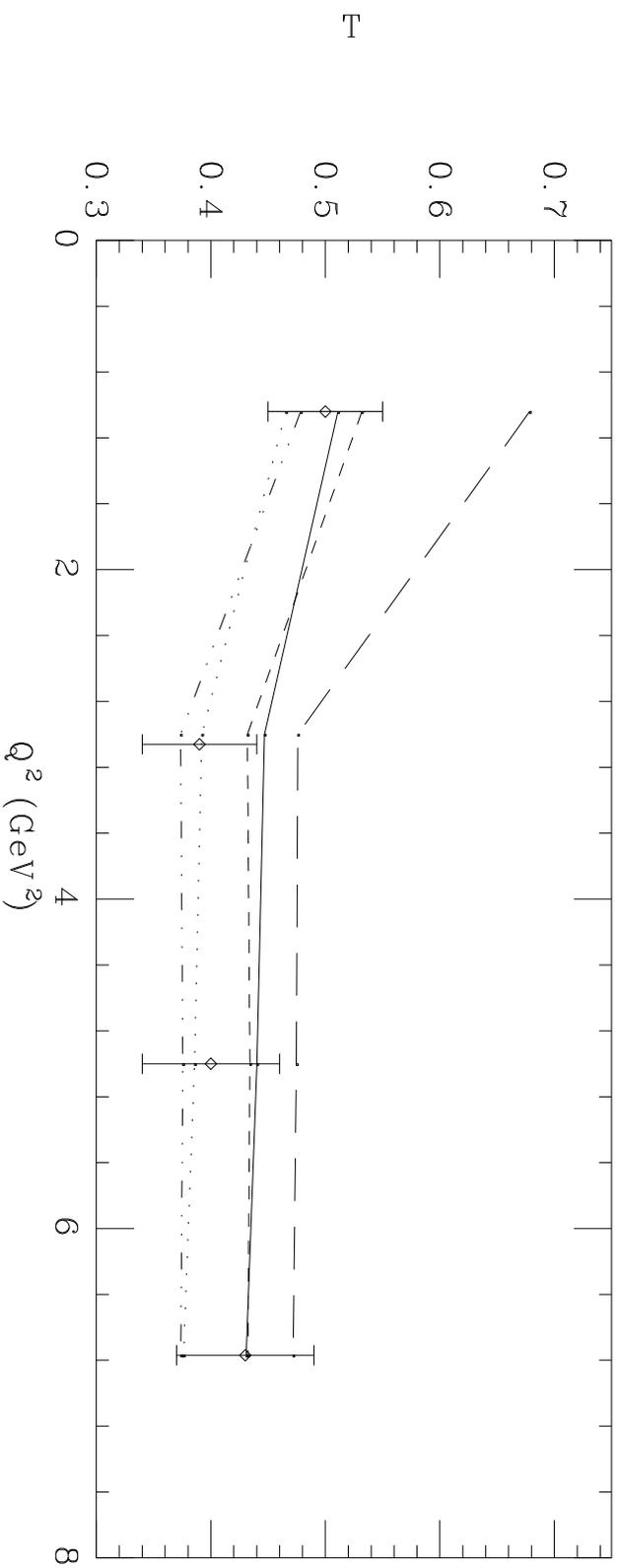
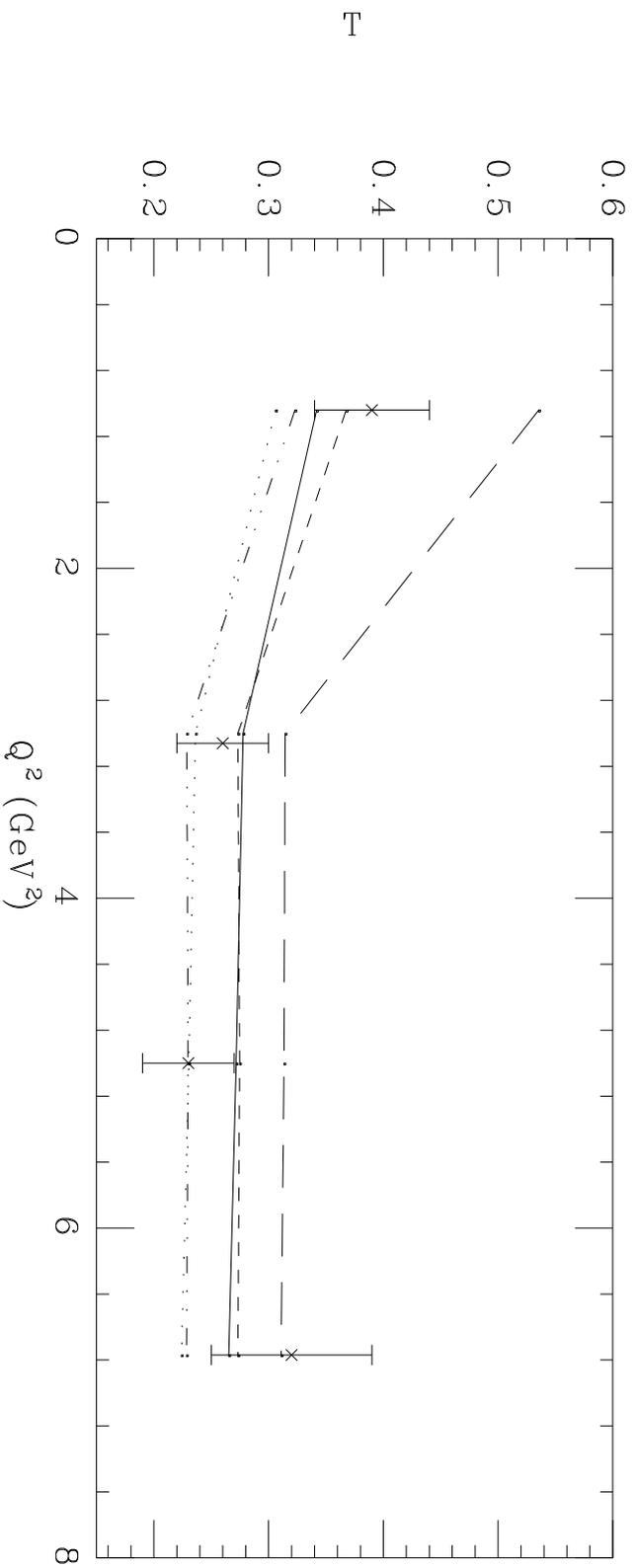
\end{document}